\documentclass[%
reprint,
amsmath,
amssymb,
aps,
prb,
floatfix,
superscriptaddress,
longbibliography,
]{revtex4-2}

\usepackage{makecell}
\usepackage{graphicx}
\usepackage[utf8]{inputenc}
\usepackage{dcolumn}
\usepackage{bm}
\usepackage[colorinlistoftodos]{todonotes}
\presetkeys{todonotes}{inline}{}
\usepackage{chemformula}
\usepackage{xspace}
\usepackage{mathtools}
\usepackage{siunitx}
\usepackage{notes2bib}
\usepackage[capitalise]{cleveref}

\newcommand{\NISO}{\ch{Ni2InSbO6}\xspace}

\newcommand{\NTO}{\ch{Ni3TeO6}\xspace}

\newcommand{\NiO}{\ch{[NiO6]}\xspace}

\newcommand{\TeO}{\ch{[TeO6]}\xspace}

\newcommand{\Niion}{\ch{Ni^{2+}}\xspace}

\newcommand{\cm}{cm$^{-1}$\xspace}
\newcommand{\TN}{$T_N$\xspace}

\begin{document}

\title{Lattice and magnetic dynamics in polar chiral incommensurate \\ antiferromagnet \NISO}

\author{M. A. Prosnikov}
 \email{mikhail.prosnikov@ru.nl}
 \altaffiliation[Currently at: ]{High Field Magnet Laboratory (HFML--EMFL), Radboud University, Toernooiveld 7, 6525 ED Nijmegen, The Netherlands}
 \affiliation{Ioffe Institute, Russian Academy of Sciences, 194021 St.-Petersburg, Russia}
\author{A. N. Smirnov}
 \affiliation{Ioffe Institute, Russian Academy of Sciences, 194021 St.-Petersburg, Russia}
\author{V. Yu. Davydov}
\affiliation{Ioffe Institute, Russian Academy of Sciences, 194021 St.-Petersburg, Russia}
\author{Y. Araki}
 \affiliation{Department of Advanced Materials Science, University of Tokyo, Japan}
\author{T. Arima}
 \affiliation{Department of Advanced Materials Science, University of Tokyo, Japan}
\author{R. V. Pisarev}
 \affiliation{Ioffe Institute, Russian Academy of Sciences, 194021 St.-Petersburg, Russia}
 
\date{\today}

\begin{abstract}
Complex systems with coexisting polarity, chirality and incommensurate magnetism are of great interest because they open new degrees of freedom in interaction between different subsystems and therefore they host a plethora of intriguing physical properties.
Here we report on optical properties and lattice and spin dynamics of \NISO single crystals studied with the use of polarized optical microscopy and micro-Raman spectroscopy in the temperature range 10--300~K. 
\NISO crystallizes in a polar structure described by the noncentrosymmetric space group $R3$ and two types of structural domains were visualized due to natural optical activity of opposite chirality.
Raman tensor elements of most $A$ and $E$ phonons along with their symmetry were determined. 
The manifestation of LO-TO splitting was observed for the $A$ modes.
By tracking the temperature dependencies of phonon frequencies the well pronounced spin-phonon interaction was observed for several modes below and above the N\'eel transition temperature \TN$=76$~K.
In antiferromagnetic phase a wide excitation centred at 247~\cm was detected and assigned to the two-magnon mode and this value was used for estimating exchange parameters through linear spin-wave theory calculations.
\end{abstract}


\maketitle

\section{\label{sec:intro}Introduction}
Breaking of the space inversion symmetry in a crystal gives rise to various intriguing phenomena.
For example, it allows the linear electrooptic effect, the piezoelectric effect, the optical second harmonic generation of the electric dipole type, etc.~\cite{sirotin_fundamentals_1982,nye_physical_1985,shen_principles_1984}.
In particular, the macroscopic electric polarization becomes allowed which strongly affects the lattice dynamics, e.g., by inducing the giant LO-TO splitting in ferroelectrics~\cite{zhong_giant_1994}, or leading to the presence of oblique phonon modes~\cite{hlinka_angular_2011}.
Another interesting property of some noncentrosymmetric crystals is the chirality, which is usually manifested as natural optical activity, and directional effects~\cite{saito_gigantic_2008,nii_microwave_2017,nomura_phonon_2019}, such as circular or linear dichroism also called as optical diode effects.
Additionally, in magnetically ordered systems the time reversal symmetry is broken which results in new degrees of freedom and possibility to control their physical properties with an external magnetic field or by varying the temperature in the vicinity of phase transitions.

Antiferromagnets with complex crystal and magnetic structures are promising candidates in the emerging fields of antiferromagnetic spintronics~\cite{gomonay_spintronics_2014,jungwirth_antiferromagnetic_2016,baltz_antiferromagnetic_2018,nemec_antiferromagnetic_2018} and sub-terahertz and terahertz magnonics~\cite{watanabe_observation_2017} due to usually richer excitation spectrum, much higher resonance frequencies and absence of stray field in comparison to ferro- and ferrimagnets.
In multiferroic and magnetoelectric crystals and artificial structures both space and time reversal symmetries are simultaneously broken pushing the interest to magnetically ordered systems further due to a coupling of multiple order parameters~\cite{pisarev_crystal_1994,pisarev_broken_1996,arima_magneto-electric_2008,dong_multiferroic_2015,cheong_broken_2018,tokura_nonreciprocal_2018,sato_nonreciprocal_2019}.

The noncentrosymmetric crystal of \NISO combines several of the above mentioned properties, namely the polarity, chirality, and incommensurate antiferromagnetic structure~\cite{ivanov_spin_2013}, making this compound an unique playground for observation of plethora of effects, excitations and their interaction.
\NISO belong to a compositionally new group of trigonal corundum-like crystals with general formula \ch{A2BB'O6}~\cite{cai_polar_2017}.
We note that this structural type is radically different from the various crystal structures realized in the so-called double perovskites for which the similar abbreviation is used~\cite{vasala_double_perovskites_2015}.
Up to now, mostly structural and static magnetic properties of the corundum-type magnetic compounds  were studied whereas their lattice and magnetic dynamics still remain practically unexplored.

Raman scattering is a well established and powerful technique allowing one to probe simultaneously lattice, electronic, magnetic, superconducting and other types of excitations as well as the coupling between them.
Additionally, polarization dependent experiments allow one to directly determine the symmetry of excitations and analyze Raman tensor elements.
It is important to remind that the mutual exclusion principle is not applicable for noncentrosymmetric systems and therefore Raman scattering allow probing of odd (polar) and even (nonpolar) phonons in a single experiment.
Recently, there was a growing interest to specific anomalies in Raman tensors in optically anisotropic media both from theoretical~\cite{kranert_raman_2016,grundmann_optically_2017,zheng_elucidation_2018} and experimental perspectives in two-dimensional materials~\cite{kim_anomalous_2015,li_revealing_2017} and bulk crystals with relatively simple structures~\cite{strach_determination_1998,kranert_Ga2O3_2016}.

In this paper, we report on a comprehensive experimental study of the optical properties, lattice and magnetic dynamics of high-quality \NISO single crystals within paramagnetic and antiferromagnetic phases.
The frequencies of polar lattice modes are the basic ingredients for the quantitative investigation of dielectric properties of the system.
Detailed lattice scattering spectra may serve as a consistent experimental reference and starting point for computations on \NISO and related crystals based on density functional theory (DFT).
We performed angular-resolved azimuthal Raman scattering measurements that allowed us to determine the Raman tensor elements and complex phases for the $A$-symmetry phonons. 
The study of magnetic excitations is essential for understanding of the ground and excited states of the spin system and, consequently, for deriving microscopic parameters, such as exchange couplings and single-ion anisotropy (SIA).
Experimental observation of a two-magnon mode was interpreted with the use of linear spin-wave theory calculations.

The paper is organized as follows.
Section~\ref{sec:exp} describes the single crystals growth process and other experimental details.
In Section~\ref{sec:results} results and their interpretation are presented, starting from the absorption spectroscopy measurements in the region of \textit{d-d} transitions in \Niion ions.
A subsection B is devoted to the optical activity study observed as a rotation of the linearly polarized light traversing the sample.
These results are followed by those of Raman scattering on lattice excitations and manifestations of the spin-phonon interaction.
Section~\ref{sec:results} ends with results on magnetic dynamics followed by the interpretation within the linear spin-wave theory approximation.
Section~\ref{sec:concl} summarizes experimental and theoretical results and ends with conclusions.

\section{\label{sec:exp}Experimental details}
Single crystals were grown by the chemical vapor transport method.
The powder samples of \NISO were synthesized by solid-phase reaction technique and put into evacuated tube with \ch{PtCl2} as a transport agent~\cite{weil_crystal_2014}.
The tube was placed for two weeks in horizontal furnace with temperatures of \SI{1080}{\celsius} and \SI{1000}{\celsius} for source and growth sides of the tube, respectively.
Grown crystals were typically of a $1\!\times\!1\!\times\!0.15$~mm$^3$ size.
They are characterized by a deep-green color specific for many crystals with \Niion ions in oxygen O$^{2-}$ octahedral coordination~\cite{lever_inorganic_1984,hugel_nature_1990,burns_mineralogical_1993,pisarev_lattice_2016,molchanova_borate_2018}.
The following notation of coordinate systems is used below: the $x$, $y$ and $z$ axes coincide with the $a$, $b$ and $c$ crystallographic axes in the hexagonal setting, while the $X$, $Y$ and $Z$ axes are used in the orthogonal setting, where $X$ and $Z$ coincide with $x$ and $z$, and $Y$ is perpendicular to the $(xz)$ plane.

Polarized optical microscopy and conoscopy measurements were done at room temperature with the use of BX53 (Olympus) microscope.
Room temperature unpolarized optical absorption spectra were recorded with the use of the UV-3600 Plus (Shimadzu) spectrometer equipped with a 0.8~mm diaphragm.

Raman spectra were measured with the use of a T64000 (HORIBA Jobin-Yvon) spectrometer equipped with a liquid-nitrogen-cooled CCD camera.
The 532~nm (2.33~eV) line of a Torus Nd:YAG-laser (Laser Quantum) was used as the excitation source.
A $50\!\times$ objective was employed for focusing the incident beam and collecting the scattered light.
Temperature-dependent spectra were recorded using a closed-cycle helium cryostat (Cryo Industries) with temperature stability better than 1~K.
Samples were mounted with the use of the silver paste for achieving good thermal contact.
Calibration of the spectrometer was done with the 520.7~\cm phonon line of a Si single crystal.
Measurements were done in the back-scattering geometry for all informative polarization settings labeled according to the Porto notation rule, where $X(YZ)\overline{X}$ stands for the case of incident light linearly polarized along the $Y$ axis, propagated and backscattered along the same $X$-axis, while the $Z$ component of scattered light is measured.
Angular azimuthal dependencies of the spectra were measured at room temperature by rotating the polished single crystal cuts of different orientations.

\section{\label{sec:results}Results and discussion}

\subsection{\label{subsec:opt_spec}Optical spectroscopy}

\begin{figure}
	\includegraphics[width=\columnwidth]{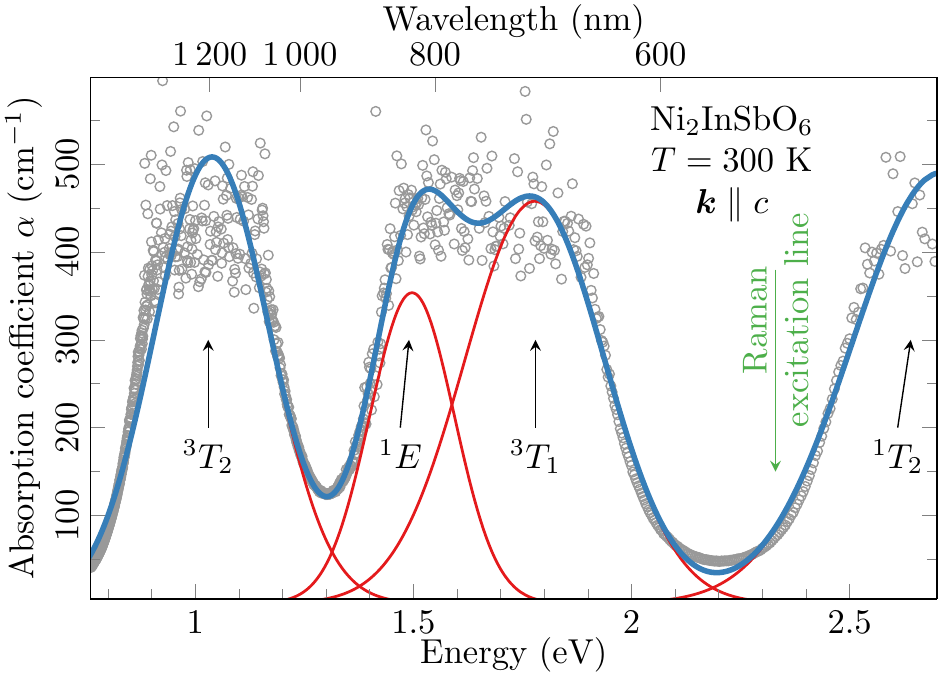}
	\caption{\label{fig:absorption} 
		(Color online) Unpolarized room-temperature absorption spectrum of \NISO for the light propagating along the optical $c$ axis.
		Observed absorption bands are due to electronic transitions withing the 3$d^{8}$ states of \Niion ions split by the cubic crystal field.  
		Experimental data are shown by gray circles; modelling of the spectra was done using the sum of Gaussian curves (red) shown in blue.
		Black arrows mark individual transitions; green arrow in the transparency region highlights the photon energy 2.33~eV (532~nm) of the excitation laser source used in Raman experiments.
	}
\end{figure}

Optical absorption spectroscopy is a widely applied technique for determination of electronic transition energies in crystal solids.
Unpolarized absorption spectrum of \NISO measured for the light propagating along the optical axis is shown in \cref{fig:absorption}.
Accuracy of the measurements in spectral regions of high optical absorption (saturation) was restricted by the thickness and small size of available samples. 
For determining the positions of the absorption bands, a simple model of four Gaussian curves was applied.
Characteristic electronic \textit{d-d} transitions from the ground $^3A_{2}$ state to the excited states of \Niion ions with energies 1.03, 1.49, 1.78 and 2.67~eV were observed.
These values are noticeably less than 1.13, 1.75, 1.95 and 2.69~eV observed in the textbook \ch{NiO} crystal composed of only cubic \NiO octahedra~\cite{newman_optical_1959,fiebig_second_2001}. 
This observation of systematically lower positions of absorption bands is an evidence of a weaker crystal field cubic parameter $Dq$ in \NISO due to the more loose local oxygen coordination of \Niion ions in comparison to \ch{NiO}.
Thus, the observed bands are assigned to $^3A_{2}$~$\rightarrow$~$^3T_{2}$, $^1E$, $^3T_{1}$, $^1T_{2}$ transitions.
Similar observations were done for \NTO~\cite{yokosuk_tracking_2015}.
The excitation source of 532~nm was chosen for Raman scattering experiments because the low absorption in the green transmittance region is favorable for maximizing the scattering volume and avoiding the resonant excitation conditions.

\subsection{\label{subsec:polarized_spec}Polarized microscopy}

\begin{figure}
	\includegraphics[width=0.95\columnwidth]{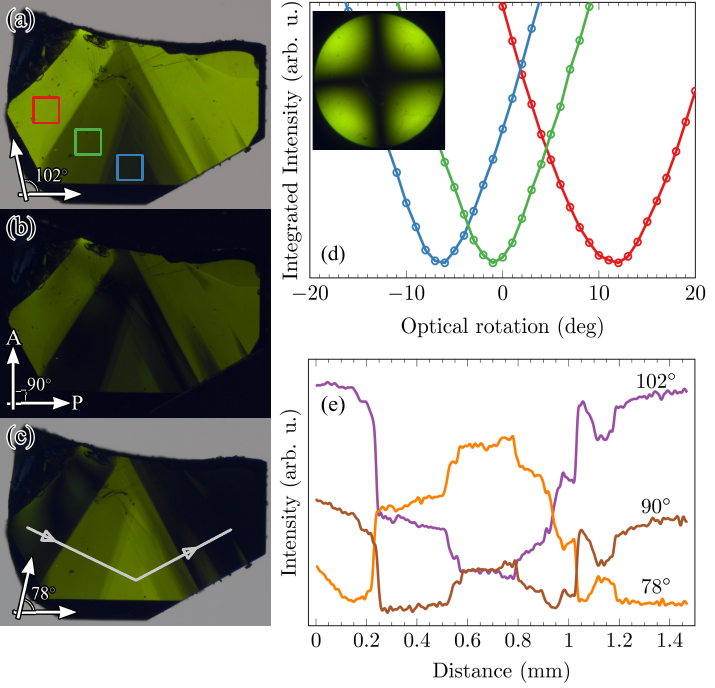}
	\caption{\label{fig:rotation} 
		(Color online) (a)-(c) Polarized microscopy images of a \NISO sample obtained for the light propagating along the optical axis and for different angles between the polarizer (P) and the analyzer (A).
		(d) Integrated intensity of the regions marked in (a) as a function of the angle difference between polarizer and analyzer.
		Inset shows the conoscopic image of the sample along the optical axis.
		(e) Transmitted intensity profiles along the gray paths shown in (c) for different polarizer/analyzer angles.
	}
\end{figure}

Polarized optical microscopy is a powerful method for fast and robust determination of different optical parameters of single crystals.
In our research, we applied this method for measuring the reciprocal effect of natural optical activity in the chiral \NISO single crystal.
This effect is described by a third rank polar tensor $\gamma_{ijk}$ invariant under time-reversal symmetry operation which relates the light vector $\bm{k}$ with the dielectric impermittivity tensor~\cite{sirotin_fundamentals_1982,aroyo_crystallography_2011}.
Due to intrinsic symmetry $\gamma_{ijk} = - \gamma_{jik}$. 
The crystal structure of \NISO is described by the point group 3 (C$_3$) and for the linearly polarized light propagating along the optical axis, $\bm{k}_3 \parallel Z$, the natural optical activity is defined by two $\gamma_{123} = -\gamma_{213}$ tensor components and is manifested as a rotation of the polarization plane of the transmitted light.

Three images of a crystal sample in the green transparency region were taken with different angles between polarizer and analyzer as shown in~\cref{fig:rotation}(a)-(c).
The characteristic dark cross observed by conoscopic measurements confirms that the sample is oriented with good accuracy along the optical axis.
These images clearly show that the sample consists of several domains with opposite chirality because the polarization plane of transmitted light is rotated in opposite directions in different areas of the sample.
Similar observations were reported for the \NTO crystal which crystallizes in the same space group $R3$~\cite{wang_interlocked_2015}.
It is interesting to note that the sample images show that domain walls are not parallel to the optical axis, but are tilted by a certain angle thus creating a zone with almost compensated rotation.

The observed sample images make it possible to determine the specific optical rotation angle of the sample with the known thickness by turning the analyzer in order to find the transmission minimum.
Such curves for selected regions are shown in~\cref{fig:rotation}(d).
The specific rotation of the linearly polarized light passing through a crystal due to optical activity is given by $\alpha = \phi/l d$, where $\phi$ is the optical rotation angle, $l$ is the sample thickness, and $d$ is the density.
For the largest rotation angle (red curve in \cref{fig:rotation}(d)) we obtain $\alpha \approx$~\SI{875}{\degree\deci\meter^{-1}\centi\meter\cubed\gram^{-1}} which is lower than the value of \SI{1355}{\degree\deci\meter^{-1}\centi\meter\cubed\gram^{-1}} determined for \NTO~\cite{wang_interlocked_2015}.
It is interesting to note that if these two notably different values are normalized to the number of Ni$^{2+}$ ions per unit cell they become equal to each other with an accuracy better than 5\%.
We note that determined value of specific rotation is comparable with the  \SI{816}{\degree\deci\meter^{-1}\centi\meter\cubed\gram^{-1}} value for the $\alpha$-quartz in the visible spectral range.

\subsection{\label{subsec:Raman_lattice}Lattice dynamics}

\begin{figure}
	\includegraphics[width=\columnwidth]{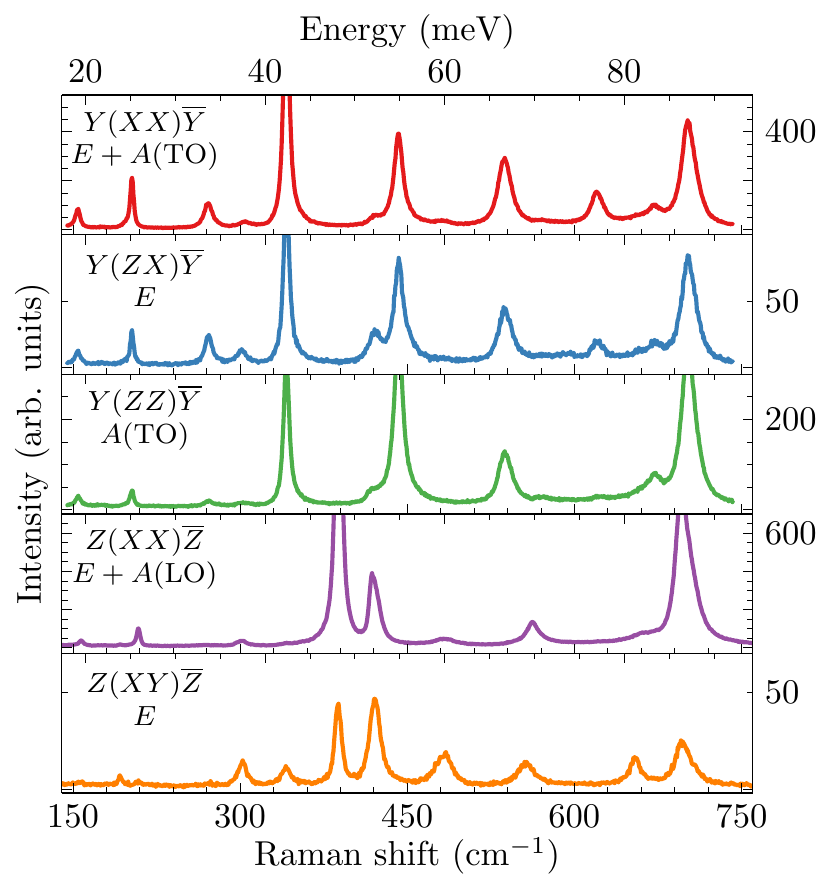}
	\caption{\label{fig:phonons} 
		(Color online) Raman spectra measured for all required polarizations at room temperature.
		Note that the ordinate axis scales are different for each spectrum.
	}
\end{figure}

\begin{figure}
	\includegraphics[width=\columnwidth]{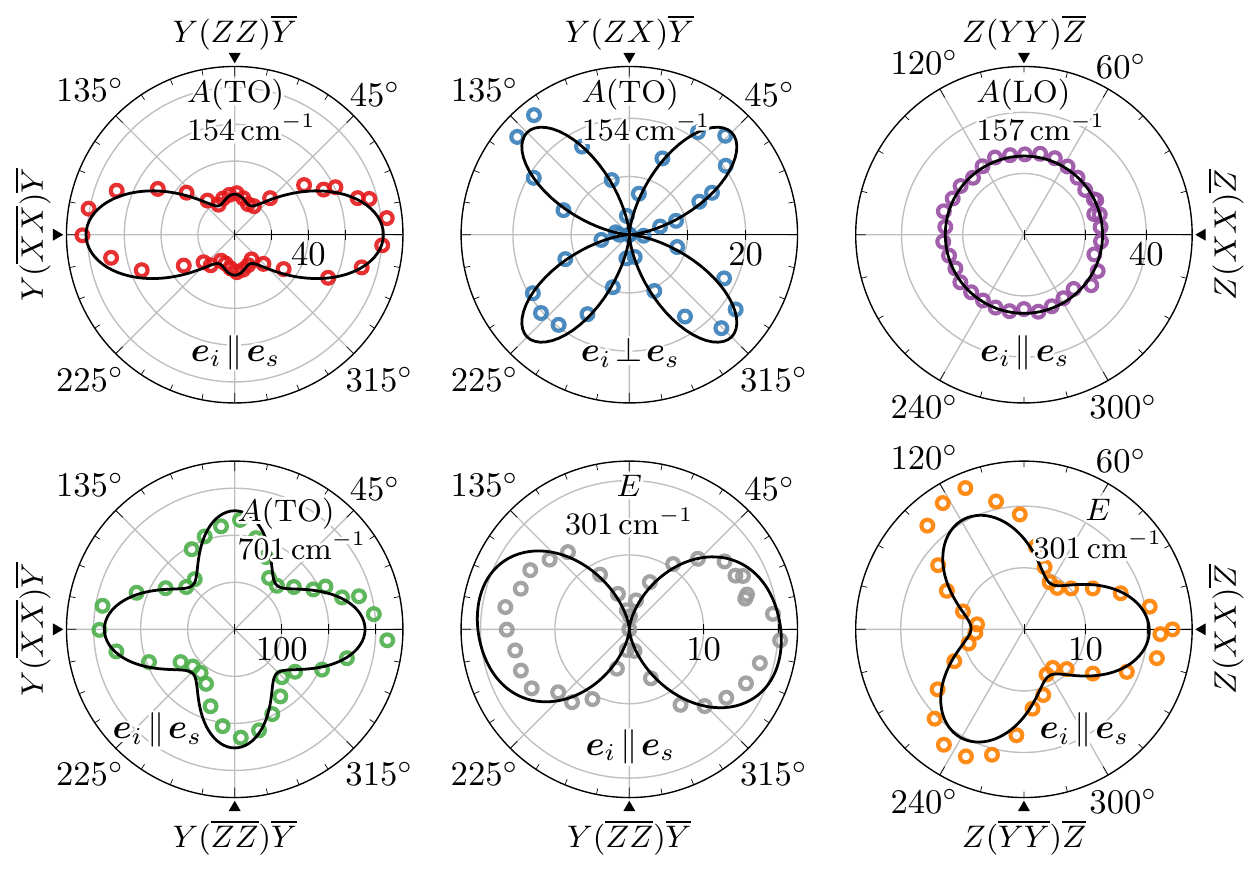}
	\caption{\label{fig:polar} 
		(Color online) Azimuthal dependencies of the intensity (arb. units) of selected phonons as a function of the polarization of the linearly polarized light.
		Colored circles represent experimental values, solid lines are fits calculated according to \cref{eq:Raman intensity}.
	}
\end{figure}


Characterization of the lattice vibrational modes is essential for a comprehensive understanding of dielectric and polar properties of crystal.
In the family of the polar corundum-like crystals the most intensively studied system is \NTO~\cite{newnham_crystal_1967,becker_reinvestigation_2006} which possesses a complex magnetic phase diagram~\cite{kim_successive_2015}.
From the point of view of magnetic properties, the important difference between \NTO and \NISO is that in the former the magnetic structure is commensurate~\cite{zivkovic_ni3teo6collinear_2010} while in the latter it is incommensurate~\cite{ivanov_spin_2013}.
Lattice dynamics of \NTO was previously characterized by infrared absorption measurements in~\cite{yokosuk_tracking_2015}, however no reports on LO-TO splitting of phonon modes in this crystal was reported.
Additionally, the lattice dynamics of only \TeO groups was studied by infrared reflectivity and Raman scattering in~\cite{blasse_vibrational_1972}.

Despite being a chiral, \NISO belongs to the achiral trigonal space group $R3$ (\#146, $Z=3$ in hexagonal cell) which was determined using the X-ray diffraction~\cite{ivanov_spin_2013,weil_crystal_2014}.
According to the group theoretical analysis 18 modes are expected which are simultaneously active in the infrared (IR) and Raman processes according to \cref{eq:modes}.

\begin{equation}
	\label{eq:modes}
		\begin{split}
			\Gamma = 9A + 9E\:.
		\end{split}
\end{equation}

In this case the $A$ and $E$ modes of Raman tensors have the following forms:

\begin{gather}
\label{eq:raman_tensors}
A =
  \begin{pmatrix*}[l]
     |a|e^{i \phi_a} & 0 & 0 \\
     0 & |a|e^{i \phi_a} & 0 \\
     0 & 0 & |b|e^{i \phi_b}
  \end{pmatrix*}, \\
E =
  \begin{pmatrix*}[l]
     \phantom{-}|d|e^{i \phi_d} & -|c|e^{i \phi_c} & -|f|e^{i \phi_f} \\
     -|c|e^{i \phi_c} & -|d|e^{i \phi_d} & \phantom{-}|g|e^{i \phi_g} \\
     -|f|e^{i \phi_f} & \phantom{-}|g|e^{i \phi_g} & \phantom{-}0
  \end{pmatrix*}.
\end{gather}

The Raman scattering intensity is given by~\cite{hayes_scattering_2012}:

\begin{gather}
	\label{eq:Raman intensity}
		I = |\bm{e}_i\,T\,\mathcal{R}\,T^T\,\bm{e}_s^T|^2,
\end{gather}

where $\bm{e}_i$ and $\bm{e}_s$ are the polarization vectors of the incident and scattered light, $T$ is the rotational matrix of the Euler angles, and $\mathcal{R}$ is the Raman tensor for the mode of particular symmetry given in \cref{eq:raman_tensors}.
In Raman scattering processes, the rotation of the polarization is equivalent to the rotation of the crystal, and the latter case was used in our experiments.
The polarization vectors were fixed such as $\bm{e}_i$ equals $(1,0,0)$ or $(0,1,0)$ in the case of parallel or crossed polarizations, respectively, while $\bm{e}_s$ was fixed as $(1,0,0)$ for all experiments.

However, the presence of a finite electric polarization along the $c$ axis allowed by the absence of the inversion center results in the LO-TO splitting of the phonon modes followed by modification of the selection rules.
Thus, the effective number of the observed modes should be greater than stated in \cref{eq:modes}.
$A$(LO) phonons should be observed only in parallel $Z(XX)\overline{Z} \leftrightarrow Z(YY)\overline{Z}$ type polarizations and characterized by a single Raman tensor element $b$ (see \cref{eq:raman_tensors}), while the $A$(TO) modes should be observed in any polarization for the light propagation direction perpendicular to the $Z$ axis~\bibnote{To include trigonal symmetry of the system, $\Theta \rightarrow \frac{2}{3} \Theta$ substitution was made for $Z(XX)\overline{Z} \leftrightarrow Z(YY)\overline{Z}$ scattering geometries}.
It should be noted that the $A$(LO) modes are inactive in the IR absorption and reflection processes, which makes Raman scattering advantageous for comprehensive characterization of the lattice dynamics.
Analysis of the Raman tensor in optically anisotropic crystals requires taking into account the complex phases between the tensor elements, and usually phase differences are found to be close to 90$^\circ$~\cite{kranert_raman_2016}.

Raman spectra of \NISO lattice modes measured at room temperature are shown in~\cref{fig:phonons}.
All observed phonon are lying in the 150--710~\cm spectral range.
The presence of rather intense second-order phonon scattering observed at frequencies above 1300~\cm additionally confirms high structural quality of the single-crystal samples.
All LO and TO phonon modes are active in the Raman scattering process and this circumstance complicates the analysis due to the strong overlapping between different modes.
Both types of the $A$ and $E$ modes were observed, however the $E$ modes are more elusive, similar to those in another model trigonal ferroelectric~\ch{LiNbO3}~\cite{margueson_e_modes_2012}.
Because of a large number of nonzero $E$ Raman tensor elements and the weakness of such phonons these elements and complex phases were determined only for the 301~\cm mode as indicated in~\cref{tab:wide_phon}.
Due to the similar reasons, the LO-TO splitting was reliably observed only for the $A$ modes.
It should be noted that obtained frequency values of the LO-TO modes can be used for estimating the dielectric permittivities $\epsilon_0$ and $\epsilon_\infty$ using the Lyddane--Sachs--Teller relation~\cite{lyddane_polar_1941} $\omega^2(\text{LO})/\omega^2(\text{TO})=\epsilon_0/\epsilon_\infty$.
The LO-TO frequency ratios for all the observed phonons are quite irregular~\bibnote{These ratios are 1.02, 1.03, 1.14, 1.04, 1.07, 1.03, and 1.01 for all the LO-TO pairs from~\cref{tab:wide_phon}}, but this is expected for real anisotropic systems.
Averaged value gives $\epsilon_0/\epsilon_\infty=1.10$.

To unambiguously determine the symmetry of phonons and extract Raman tensor elements we performed the measurements by varying the azimuthal angle of crystal with respect to linearly polarized incident light.
Spectra were fitted with Voigt profiles and linear background and typically resulting values of $R^2$ were $\geq\!0.98$.
For each observed phonon the fitting was done simultaneously for all available polarizations, e.g., (a) and (b), or (e) and (f) plots in \cref{fig:rotation}.
This approach allowed us to extract individual tensor elements and phases.
Necessity to use complex phases in the case of the \NISO crystal is directly confirmed by nonvanishing intensity of the $A$(TO) modes when measured at 45$^\circ$ to the $X$ (or $Y$) axis, as can be seen for the  154~\cm mode in \cref{fig:polar}.
All complex phases for the $A$ modes were close to expected 90$^\circ$ value.
The deviations from this value was mainly observed in few- and single-monolayer thickness samples~\cite{li_revealing_2017}.
Estimated Raman tensor elements along with phase differences are summarized in \cref{tab:wide_phon}.
We should note that the absolute values of tensor elements are instrument-dependent, so for comparison with other experiments $a/b$ ratio should be used.

No frequency shifts more than 2~\cm were observed in angular-resolved measurements which rules out the presence of oblique modes observed for, example, in \ch{BiFeO3}~\cite{hlinka_angular_2011}.
We suppose that the absence or smallness of the frequency shifts could be due to a small value of the electric polarization \SI{6}{\micro \coulomb \per \centi \meter \squared} of \NISO~\cite{weil_crystal_2014} in comparison with other polar systems where angular dispersion shifts were clearly observed~\cite{otaguro_phonons_1971,hartwig_phonon_1971}.

\begin{table*}
	\caption{\label{tab:wide_phon}
		Frequency (\cm), FWHM (\cm), Raman tensor elements, phase difference (deg.) and spin-phonon coupling constant (\cm) for the observed phonon modes.
		Constants $\lambda$ are determined from temperature measurements, other parameters are given for the room temperature.
	}
	\begin{ruledtabular}
		\begin{tabular}{c c|c c|c c|c|c|cc||c|c|c}
            \multicolumn{10}{c||}{$A$(TO) ($A$(LO))} & \multicolumn{3}{c}{$E$}\\
 			\hline
            \multicolumn{2}{c|}{$\omega$} & \multicolumn{2}{c|}{FWHM} & \multicolumn{2}{c|}{a} & b & $\phi_{a-b}$ & \multicolumn{2}{c||}{$\lambda$} & $\omega$ & FWHM & $\lambda$\\
			\hline
			155 & (158)   & 6.1 & (5.4)     & 8.99      & (5.07)      & 4.69    & 87.7   & --- & --- & 192  & 5.6  & --- \\
			204 & (210)   & 4.1 & (3.9)     & 14.89     & (9.68)      & 6.04    & 88.8   & --- & --- & ---  & ---  & --- \\
			342 & (389)   & 6.2 & (6.6)     & 33.74     & (43.24)     & 19.01   & 88.8   & $-0.55$ & $(-1.78)$ & 271  & 9.6  & --- \\
			--- & (419)   & --- & (5.5)     & ---       & (15.08)     & ---     & ---    & --- & $(-2.14)$ & 302\footnote{The values used to fit this mode are: $c = 3.8, d = 3.7, e = 3.8, f = 3.4, \phi_d = 45.8^\circ, \phi_e = -20.2^\circ, \phi_e = 18.9^\circ, \phi_f = -51.9^\circ$}  & ---   & ---  \\
			444 & ( --- ) & 11.2 & ( --- )  & 19.56     & ( --- )     & 20.53   & 88.8   & --- & --- & 351  & 17.6  & --- \\
			539 & (563)   & 14.6 & (16.7)   & 16.95     & (10.46)     & 11.11   & 90.5   & --- & --- & 423  & 7.4  & --- \\
			621 & (662)   & 11.9 & (20.7)   & 11.76     & (4.21)      & 2.80    & 80.2   & --- & --- & 484  & 14.3  & $-0.68$ \\
			673 & (696)   & 18.1 & (11.6)   & 8.44      & (25.28)     & 7.52    & 85.4   & --- & --- & ---  & ---  & --- \\
			702 & (706)   & 12.4 & (17.9)   & 16.65     & (16.98)     & 15.88   & 91.1   & $-1.13$ & $(-0.82)$ & ---  & ---  & --- \\
		\end{tabular}
	\end{ruledtabular}
\end{table*}

\subsection{\label{subsec:spin_phonon} Spin-phonon coupling}

\begin{figure}
	\includegraphics[width=\columnwidth]{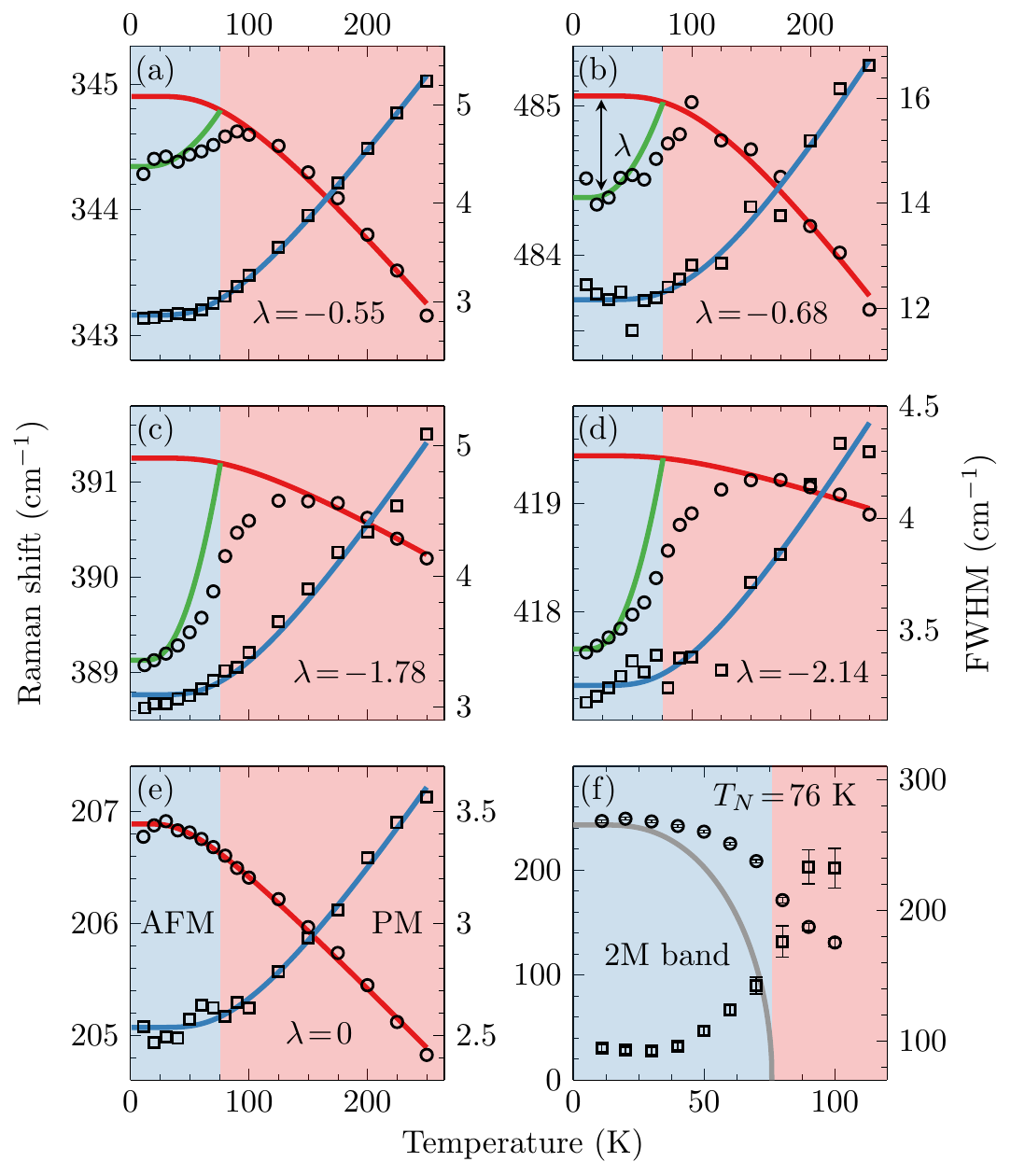}
	\caption{\label{fig:spin_phonon} 
		(Color online) Temperature dependencies of frequency (circles) and FWHM (squares) of (a)-(e) phonon and (f) two-magnon excitations.
		Blue and red regions correspond to antiferromagnetic and paramagnetic phases.
		Red and blue curves are anharmonic contributions according to \cref{eq:freq,eq:FWHM}.
		Green curves are spin-phonon contribution.
		Gray curve is the Brillouin function for $S=1$.
		}
\end{figure}

Temperature dependent experiments revealed well pronounced frequency shifts of some phonon modes above and below \TN.
Temperature dependencies of the frequency and half-width of several particular phonons are shown in~\cref{fig:spin_phonon}.
Typically, the conventional behavior of phonon modes with cooling is characterized by simultaneous narrowing and hardening due to thermal suppression of anharmonism, see, e.g.,~\cref{fig:spin_phonon}(e).
The anharmonic curves for both frequency and half-width in \NISO were calculated with the use of the well known equations~\cite{balkanski_anharmonic_1983}.
For explaining the observed behavior the three phonon relaxation process was found to be sufficient.
This process is subject to the following theoretical predictions for the frequency and half-width: 

\begin{gather}
	\label{eq:freq}
	\omega_i(T) = \omega_{i0} - A \left( 1 + \frac{2}{e^{\hbar \omega_{i0}/2k_BT} - 1} \right),
\end{gather}

\begin{gather}
	\label{eq:FWHM}
		\Gamma_i(T) = \Gamma_{i0} + C \left( 1 + \frac{2}{e^{\hbar \omega_{i0}/2k_BT} - 1} \right),
\end{gather}

where $\omega_{i0}$ and $\Gamma_{i0}$ are quasi-harmonic frequency and fullwidth (instrument limited), respectively.
$k_B$ is the Boltzmann constant, $T$ represent temperature, $A$ and $C$ are anharmonic constants 

Part of the phonons (plots (a)-(d) in~\cref{fig:spin_phonon}) reveals anomalous frequency shifts while FWHM's were well described by anharmonic curves thus obeying characteristic behavior in the case of spin-phonon interaction~\bibnote{The anharmonic constants obtained through a fitting procedure are: $\omega_0 = 346.31, A = -1.41, \Gamma_0 = 0.77, C = 2.08$; $\omega_0 = 487.11, A = -2.05, \Gamma_0 = 5.15, C = 7.01$; $\omega_0 = 392.31, A = -1.06, \Gamma_0 = 1.08, C = 2.02$; $\omega_0 = 420.01, A = -0.57, \Gamma_0 = 1.88, C = 1.37$; $\omega_0 = 207.70, A = -0.82, \Gamma_0 = 1.28, C = 1.26$ for curves (a)-(e) in~\cref{fig:spin_phonon}, respectively.}
It is also curious to note that all the modes subject to such interaction undergo softening (negative spin-phonon coupling constants).
Another part of phonons (plots (c) and (d)) manifest strong shifts by 2~\cm which is almost twice as large in comparison to hardening in the 10--300~K range due to anharmonicity.
Such strong spin-phonon coupling is not typical for small-spin ($S=1$) systems for \Niion~\cite{prosnikov_NiW_2017,prosnikov_NiNb_2018}, however comparable and even larger shifts were observed in \NTO~\cite{yokosuk_tracking_2015}.

Experimental values of spin-phonon constants $\lambda$ are determined as a frequency difference of the phonon mode at the lowest temperature (10~K) and that predicted by anharmonic equation (\cref{eq:freq}) and listed in \cref{tab:wide_phon}.
In the case of the $S=1$ ions this difference equals to the spin-phonon constant $\lambda$.
The contribution of spin-phonon interaction to phonon frequencies can be described by a static averaged spin-spin correlation function $\lambda \langle \mathbf{S}_i \cdot \mathbf{S}_j \rangle$ for adjacent spins~\cite{lockwood_spin-phonon_1988}.
Thus, this contribution within the mean field approach is expressed as $(\langle S^z \rangle / S)^2$, where $\langle S^z \rangle$ is the Brillouin function.
Fitting of the experimental data for \NISO gives satisfactory results with spin-phonon coupling constants $\lambda$ different (or equal to zero) for different phonons as shown in \cref{fig:spin_phonon}.
However, visible discrepancies of spin-phonon contributions with experimental values in the vicinity and above the N\'eel temperature are seen which points to important contribution of the short-range magnetic ordering.
The existence of such short-range correlations in the paramagnetic phase is independently confirmed by the presence of the two-magnon mode well above \TN (see \cref{fig:spin_phonon}(f)).

\subsection{\label{subsec:Raman_spin} Magnetic dynamics}

\begin{figure}
	\includegraphics[width=\columnwidth]{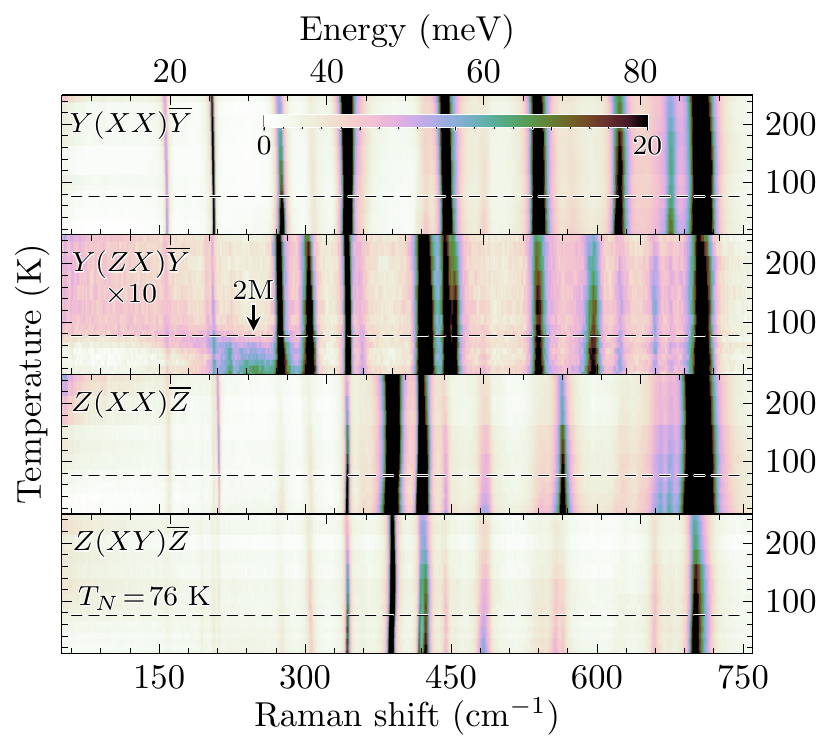}
	\caption{\label{fig:map} 
		(Color online) Temperature maps of Raman spectra.
		Dashed horizontal line indicates the temperature of the antiferromagnetic phase transition at \TN=76~K.
		Note different scale for $Y(ZX)\overline{Y}$ map.
	}
\end{figure}

At low temperatures a pronounced wide band with maximum centered at 247~\cm was observed in the scattering spectra, see $Y(ZX)\overline{Y}$ spectrum in \cref{fig:map}.
This excitation was observed in $Y(ZX)\overline{Y}$, $Z(XY)\overline{Z}$, and $Z(XX)\overline{Z}$ polarizations with absolute intensities ratio of 1.3:1.2:0.5.
The band is strongly dampens and the frequency is soften while approaching the N\'eel temperature, as shown in~\cref{fig:spin_phonon}(f).
However, it can be observed even above \TN  as a broad quasi-elastic wing up to 150~K.
Temperature dependence, overall half-width, and energy range suggest two-magnon origin of this mode, which is a process of simultaneous excitation of magnon pair with wavevectors $\bm{k}$ and $-\bm{k}$~\cite{fleury_scattering_1968}.
No other excitations of magnetic nature were observed in other spectra.

\subsection{\label{subsec:LSWT} Linear spin-wave theory calculations}

\begin{figure*}
	\includegraphics[width=17.8cm]{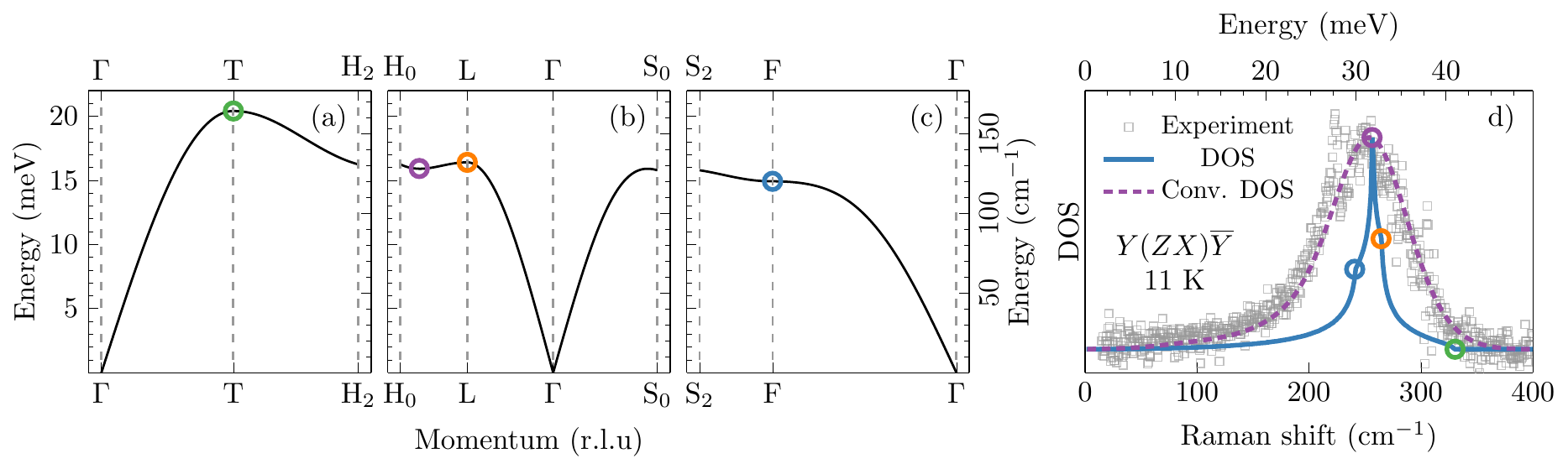}
	\caption{\label{fig:disp} 
		(Color online) (a)-(c) Spin-wave dispersion for the high-symmetry directions in rhombohedral setting.
		(d) Two-magnon Raman spectrum at $T$=11~K (gray squares) shown with removed phonon background contribution. 
		Magnon DOS (red curve) and DOS convoluted with Gaussian (dashed curve), see text for details.
		Note that energy of DOS curves is doubled to match the two-magnon process. 
		Circles highlight Van Hove singularities corresponding to particular points in the dispersion curves.
	}
\end{figure*}

It is known that the magnetic structure of \NISO is antiferromagnetic and incommensurate, however the magnetic propagation vector $\bm{k}=(0,0.036(1),0)$ is rather small~\cite{ivanov_spin_2013} and the smallness of this value allows us to do calculations accepting the following assumptions:

i) The AFM structure is supposed to be commensurate with $\bm{k}=0$.

ii) We will assume that $J1=J2$ and $J3=J4$ which is strictly not the case, because the bond lengths for $J1$ and $J2$ are different, 3.76343 and 3.86013~\AA, respectively, according to structural measurements~\cite{ivanov_spin_2013}).

For $J3$ and $J4$ exchange couplings bond lengths are equal, however they might be different because they are not symmetry related.

For testing the hypothesis of the two-magnon nature of the observed 247~\cm mode the dispersion curves were numerically calculated within the linear spin-wave theory approach~\cite{toth_linear_2015} and, subsequently they were used for the density of states (DOS) calculations. 
Note that dispersion curves and points in~\cref{fig:disp} are shown for the primitive (rhombohedral) basis.
Calculations predict one doubly degenerate acoustic magnon mode, as expected, since there are two magnetic ions within primitive cell.

As the first approach, only $J1\!=\!J2\!=\!2.7$~meV exchange constants were taken into account which resulted in a three-dimensional magnetic structure with the maximum of DOS compatible with the position of the observed two-magnon band.
Since there are no optical magnons to constrain estimation of exchange constants we may use paramagnetic Curie-Weiss temperature $\Theta = -184$~K, measured in~\cite{ivanov_spin_2013}.
This value represents averaged exchange couplings and for \NISO it is expressed as

\begin{gather}
	\label{eq:Theta}
		\Theta = - \frac{S(S+1)}{3 k_B} \sum_i z_i J_i\:, 
\end{gather}

where $S$ is the spin number, $z$ is the number of neighboring ions, and summarizing goes up for all exchange couplings.
For the case of $J1\!=\!J2\!=\!2.7$~meV it gives us the value $\Theta = -125$~K which is significantly different from $\Theta = -184$~K.
This discrepancy between experiment and calculation suggests that the next-neighbor exchange couplings have to be taken into account.
It should be additionally noted that this approximation gives completely non-frustrated magnetic structure which contradicts experimental observation since the frustration index is $|\Theta/N|=2.4$~\cite{ivanov_spin_2013}.
This value is not big enough for considering \NISO as a frustrated magnet, but suggests the presence of non-negligible frustrations. 

Now we include additional next-neighbour exchange paths $J3\!=\!J4$ connecting ions within the $XY$ planes with Ni--Ni distances of 5.2164~\AA.
Though these exchange paths connect ferromagnetically aligned spins they were set to be antifferomagnetic in order to introduce small frustration and further decrease of $\Theta$.
Such a model results in $J1\!=\!J2\!=\!3.4$ and $J3\!=\!J4\!=\!0.5$~meV and gives $\Theta = -181$~K which is in the good agreement with the experimental value.

The dispersion curves along the particular paths calculated with the use of \textsc{SeeK-path} code~\cite{hinuma_band_2017} and are shown in~\cref{fig:disp}~(a)-(c).
Since the frequency of acoustic magnon branch is below the range achievable in our experiments we will base our calculations on the observed two-magnon band whose position and shape should be proportional to the density of magnon states.
However, it is known that in the process of the second order scattering the magnon-magnon interaction is important~\cite{elliott_effects_1969} which results in strong damping making singular points in the density of states indistinguishable.
To simulate damping effects, the density of states was convoluted with Gaussian profile of 4~meV fullwidth.
Good agreement of the calculated density of states and experiment is shown in~\cref{fig:disp}~(d).

No other signs of magnetic scattering were observed down to $15~\text{cm}^{-1}\!\approx\!2$~meV which allows us to suppose that the acoustic magnon branch has energy below this value and, consequently, the single ion anisotropy is small.
In fact, for the proposed exchange model the estimated value of anisotropy energy results in a value $-0.05$~meV for the case when the easy axis is directed along the $Z$ axis.
This value of the anisotropy results in a gap of $\approx$2~meV in the spin-wave spectrum at the Gamma point.
Further progress in the understanding magnetic dynamics of \NISO crystals can be reached in the studies of the antiferromagnetic resonance with the use of microwave technique or direct measurement of spin-wave dispersion by $\bm{k}$-resolved techniques, such as inelastic neutron or X-ray scattering.

\section{\label{sec:concl} Summary and conclusions}
In summary, we report on comprehensive study of a polar chiral incommensurate antiferromagnet \NISO with the use of absorption spectroscopy, optical microscopy, and polarized Raman spectroscopy.
This system crystallizes in noncentrosymmetric space group $R3$ and undergoes a phase transition at \TN$=76$~K into incommensurate antiferromagnetic phase.
Room temperature optical absorption spectroscopy revealed four electronic excitations with energies 1.03, 1.49, 1.78 and 2.67~eV which we assigned to $^3A_{2}$~$\rightarrow$~$^3T_{2}$, $^1E$, $^3T_{1}$, $^1T_{2}$ electronic transitions within the 3$d^8$ states of \Niion ions split by the crystal field.
Optical microscopy allowed us to observe two types of chiral domains in a single crystal sample which are related to each other by inversion with the natural optical activity of opposite sign. 
Most phonons predicted by the group-theoretical analysis, including the split LO-TO phonons, were observed.
Raman tensor elements and phase differences were determined for the $A$-symmetry modes by performing angular-resolved azimuthal polarization measurements.
Temperature measurements revealed pronounced spin-phonon interaction manifested as softening of particular modes below and slightly above \TN due to short-range magnetic ordering.
A wide asymmetric two-magnon band with a maximum of 247~\cm ($T=11$~K) was observed in Raman spectra with $Y(ZX)\overline{Y}$, $Z(XY)\overline{Z}$, and $Z(XX)\overline{Z}$ polarizations.
Based on the observed two magnon excitation and known paramagnetic Curie-Weiss temperature
we propose an exchange model with parameters $J1\!=\!J2\!=\!3.4$ and $J3\!=\!J4\!=\!0.5$~meV which was used for calculating dispersion curves and density of states within linear spin-theory.
Our results contribute to a better understanding of lattice and magnetic dynamics in polar magnets with corundum-type structures and the developed model of exchange structure and spin-wave dispersion curves can be directly applied to other similar systems.

\begin{acknowledgments}
This work was funded by the Russian Science Foundation according to the research project No.\,16-12-10456.
\end{acknowledgments}

\bibliography{niso}

\end{document}